\documentclass[twocolumn, usenatbib]{mnras}
\usepackage{savesym}
\usepackage{graphicx}
\usepackage{longtable}
\usepackage{changepage}
\usepackage{diagbox}

\usepackage{subfig}
\usepackage{amsmath}
\usepackage{amssymb}
\usepackage{verbatim}
\usepackage[yyyymmdd,hhmmss]{datetime}
\usepackage{array}
\usepackage{times}
\usepackage{xcolor}

\DeclareRobustCommand{\DE}[3]{#2}
\let\DEthebibliography\thebibliography
\def\thebibliography{\DeclareRobustCommand{\DE}[3]{##3}\DEthebibliography}

\newcommand\met{{\cite{Mummery24Plunge}}}

\newcommand{\nustar}{\textit{NuSTAR}}

\newcommand{\swift}{{\it Swift}}

\newcommand{\nicer}{{\it NICER}}

\newcommand{\src}{{MAXI J0637$-$430}}
\newcommand\fk{{{\tt fullkerr}\ }}

\title[Plunging region emission in MAXI J0637$-$430]{ Plunging region emission in the X-ray binary MAXI J0637$-$430 }
\author [Andrew Mummery, et al.]{Andrew Mummery$^1$\thanks{E-mail:
andrew.mummery@physics.ox.ac.uk}, Jiachen Jiang$^2$, Andrew Fabian$^2$ 
\\
$^1$Oxford Theoretical Physics, Beecroft Building,  Clarendon Laboratory, Parks Road, Oxford, OX1 3PU, United Kingdom \\
$^2$Institute of Astronomy, Madingley Road, Cambridge CB3 0HA
}

\date{}

\begin{document}

\pagerange{\pageref{firstpage}--\pageref{lastpage}} \pubyear{2023}

\maketitle

\label{firstpage}

\begin{abstract} 
On the second of November 2019 the black hole X-ray binary MAXI J0637$-$430 went into outburst, at the start of which it was observed in a thermal ``disc-dominated'' state. High photon energy (extending above 10 keV) observations taken by the \nustar\ telescope reveal that this thermal spectrum can not be fit by conventional two-component (disc plus corona) approaches which ignore disc emission sourced from within the plunging region of the black hole's spacetime. Instead, these models require a third ``additional'' thermal component to reproduce the data. Using new disc solutions which extend classical models into the plunging region we show that this ``additional'' thermal emission can be explained self-consistently with photons emitted from the accretion flow at radii within the innermost stable circular orbit of the black hole.   This represents the second low mass X-ray binary, after MAXI J1820+070, with a detection of plunging region emission, suggesting that signatures of this highly relativistic region may well be widespread but not previously widely appreciated. {To allow for a detection of the plunging region, the black hole in MAXI J0637$-$430 must be at most moderately spinning, and we constrain the spin to be $a_\bullet < 0.86$ at 99.9$\%$ confidence. } We finish by discussing the observational requirements for the robust detection of this region. 
\end{abstract}

\noindent


\section{Introduction} 
The discovery and analysis of astrophysical black holes through the detection of X-ray photons emitted from deep within their spacetimes is by now a well established field in high energy astrophysics. As the accretion flows which form in black hole binary systems reach temperatures of order a few kiloelectronvolts in their inner regions,  X-ray photons observed from galactic binaries have, in principle, the characteristic signatures of  highly relativistic regions of spacetime imprinted onto their spectral energy distributions (SEDs).  The set of observational techniques employed in analysing these broadband SEDs with the intention of inferring black hole properties is  named “continuum fitting”, and has been employed widely throughout the literature \citep[e.g.,][]{Shafee06, Steiner09, McClintock14, Zhao21}.  

In the context of continuum fitting one particularly interesting, and relevant,  feature of black hole spacetimes is the existence of a so-called ISCO, or innermost stable circular orbit, radius. A purely relativistic effect, the ISCO radius represents the innermost point in a black hole's spacetime at which circular orbits remain stable to inwards perturbations. Interior to the ISCO radius any inward perturbation of a fluid element will pick up a rapidly accelerating radial velocity, and will ultimately cross the event horizon of the black hole. Whether or not substantial (or even observable) photon fluxes are sourced within this plunging region has been a long controversial theoretical question, discussed in some of the very first papers on black hole accretion theory \citep{PageThorne74}. 

The temperature of black hole accretion flows were, in the early days of thin disc theories, expected to drop to $\sim$ zero within the plunging region as a purely viscous stress would be unable to transport angular momentum across this boundary and therefore no dissipation would occur. However, with the realisation that it is in fact magnetohydrodynamic turbulence that redistributes angular momentum in an accretion flow \citep{BalbusHawley91}, which can in principle transfer angular momentum across this boundary \citep{Gammie99, Krolik99, AgolKrolik00}, the question of the ISCO temperature was reopened.   Numerical general relativistic magnetohydrodynamical simulations \citep[e.g.,][]{Noble10, Penna10, Zhu12, Schnittman16, Lancova19, Wielgus22} generically find non-zero temperatures at and within the ISCO, with non-zero associated thermal emission, suggesting that magentic stresses do indeed continue to dissipate in this region. The vanishing ISCO temperature boundary condition has however retained prominent advocates \citep[e.g.,][]{Paczynski00}, although these precise arguments are in contention with simulations \citep[e.g.][]{Noble10}. 


It is only recently, with the development of analytical models for the thermodynamics of this plunging region \cite{MummeryBalbus2023, MummeryMori24}, that observational data (through the medium of continuum fitting) can be leveraged to probe the question of whether or not observable photon fluxes are sourced within the plunging region. \met\ recently demonstrated that the thermal X-ray spectrum of the black hole X-ray binary MAXI J1820+070 could only be adequately described with the inclusion of photons emitted from within the plunging region, while traditional models produced insufficient flux at $E \sim 6-10$ keV \citep{Fabian20}. Other sources, such as MAXI J1828$-$249, have also been observed to show similar features in one-off observations \citep{Oda19}. These results open up the possibility of probing this highly relativistic regime with observations of Galactic X-ray binaries. 


On the second of November 2019 the black hole X-ray binary MAXI J0637$-$430 went into outburst \citep{Negoro19,Tetarenko21}, at the start of which it was observed in a thermal ``disc-dominated'' state \citep{Lazar21}.  High photon energy (extending above 10 keV) observations taken by the \nustar\ telescope reveal that this thermal spectrum can not be fit by conventional two-component (disc plus corona) approaches which do not include disc emission sourced from within the plunging region of the black hole's spacetime \citep{Lazar21}. 
Instead, these models require a third {\it ad-hoc} ``additional thermal component'' to be added in by hand to reproduce the data. This behaviour is identical to that originally observed in the X-ray binary MAXI J1820+070 \citep{Fabian20}. It is the purpose of this {\it Letter} to analyse three  high quality X-ray spectra taken of MAXI J0637$-$430, with the aim of probing the physics of the plunging region in this source. 
We will demonstrate that a satisfactory fit of the MAXI J0637$-$430 data requires intra-ISCO emission at high significance, with the flux sourced from within the ISCO representing $\sim 40\%$ of the total observed flux at $E \sim 5-8$ keV.  

\section{Analysis}
\subsection{The disc model}
For this analysis we use the {\tt fullkerr} \met\ model in {\tt XSPEC} \citep{Arnaud96}, which smoothly joins the steady state relativistic thin disc solutions of \cite{NovikovThorne73} onto the thermodynamic solutions valid for the intra-ISCO plunge derived by \cite{MummeryBalbus2023, MummeryMori24}. This model takes as input the usual free parameters of a thin-disc description of a black hole accretion flow. The black hole is described by it's mass $M_\bullet$, and spin $a_\bullet$, while the disc has a constant accretion rate $\dot M$ and is located a distance $D$ from the observer with normal orientated at an angle $i$ from the line of sight. This accretion flow is further assumed to evolve in the equatorial plane of the black hole, making the disc normal equal to the black hole's spin axis. {In the main body of the disc (i.e., at radii much larger than the ISCO) the} {\tt fullkerr} {model is functionally identical to the familiar} {\tt kerrbb} {model.} To smoothly join {this outer model} onto a plunging intra-ISCO flow an ISCO stress parameter $\delta_{\cal J}$ is included, which {sets the ISCO temperature boundary condition, and physically} parameterises the fraction of the disc fluids angular momentum which is ``passed back'' to the extra-ISCO accretion flow. {Numerical simulations} \citep[e.g.,][]{Noble10, Penna10, Zhu12, Schnittman16} {typically place $\delta_{\cal J}$ in the region of $\delta_{\cal J} \simeq 0.01-0.1$} \citep[see][for a detailed discussion of the physics of $\delta_{\cal J}$]{MummeryBalbus2023, Mummery24Plunge}. 

The physics of radiative transfer in the main (extra-ISCO) body of the disc is parameterised with a simple colour-correction factor $f_d$, while within the ISCO accretion flows are known to transition to a photon-starved state {where the disc fluid is unable to produce the required number of photons to carry away the locally radiated disc luminosity with a blackbody spectrum} \citep{Zhu12, Davis19}. We model the physics of this transition with a tuneable parameter $\xi$, which is discussed in \met. {Briefly, the colour-correction factor grows within the ISCO as $f = f_d (r_I/r)^\xi$, which accounts for the increased energy each photon must carry in a photon-starved state.} In principle therefore there are 8 free parameters which could be fit to the thermal continuum observed from MAXI J0637$-$430: $M_\bullet, a_\bullet, \dot M, D, i, \delta_{\cal J}, f_d$ and $\xi$. However, some of these parameters are somewhat constrained by prior information, as we now discuss. 

\subsection{Prior information regarding MAXI J0637$-$430 properties}
The X-ray binary MAXI J0637$-$430 has been  widely studied in the literature \citep[e.g.,][]{Tetarenko21, Jana21, Lazar21, Baby21, Thomas22, Ma22, Soria22, Draghis2023, Jia23}. The results of many of these analyses are somewhat contradictory, but there are broad regions of physical parameter space which do overlap between works. The compact object at the centre of MAXI J0637$-$430 has, in a number of works, been reported to be a black hole \citep[e.g.,][]{Jana21, Lazar21, Soria22}, with mass too high to be a neutron star. The mass of the compact object is typically inferred to be in the range $M_\bullet/M_\odot \sim 4-12$, with the smallest uncertainty quoted in \cite{Soria22} at $M_\bullet = (5.1 \pm 1.6) M_\odot$. The black hole spin quoted in different works however varies substantially, \cite{Baby21} favour a retrograde $a_\bullet < 0$ spin, and a low spin is also favoured by \cite{Soria22} $a_\bullet \lesssim 0.25$. 
High spins are however favoured by \cite{Jia23} $a_\bullet > 0.75$ and \cite{Draghis2023} $a_\bullet = 0.97 \pm 0.02$. 
As black hole spin measurements are typically somewhat degenerate with a change in disc-observer inclination, different works also favour different inclination angles between the observer's line of sight and the black hole's spin axis. Most works broadly agree that MAXI J0637$-$430 is located at a typical disc-observer distance of $D \sim 8$ kpc, with \cite{Soria22} finding $D = 8.7 \pm 2.3$ kpc. 

As there is no agreed upon set of fundamental parameters which describe the MAXI J0637$-$430 system, we will limit the analysis in this work to seeking to identify and analyse the signatures of the plunging region, while fixing the other physical parameters of the system to plausible values.

\subsection{The data}
We are interested in high quality observations of MAXI J0637$-$430 in the soft disc-dominated state, and require sufficient high photon energy coverage to disentangle the disc and Comptonised components. \cite{Lazar21} demonstrated that the detection of photons with energies $E \gtrsim 10$ keV are required to differentiate between different spectral models.  As such we focus our attention on three \nustar\ observations which probe these higher energy ranges $(E = 3-78$ keV$)$. To ensure we have a good handle on the properties of the disc we also include contemporaneous observations taken with \nicer\ and \swift\ XRT, which probe lower energies closer to the peak temperature of the accretion flow $(E \simeq 0.5-10$ keV$)$. The three \nustar\ observations were taken on the 5$^{\rm th}$, 14$^{\rm th}$ and $25^{\rm th}$ of November 2019 (58792, 58801 and 58812 MJD). The \nicer\ observation was taken on 58792 MJD, and the \swift\ observations were taken on 58801 and 58812 MJD. The observational details are summarised in Table \ref{tab_obs}.


\begin{table*}
    \centering
    \begin{tabular}{ccccccc}
    \hline\hline
    MJDs & Missions & ObsIDs & Length (ks) & Missions & ObsIDs & Length (ks) \\
    \hline
    58792.6  & \nicer\ & 2200950103 & 0.6 & \nustar\ & 80502324002 & 36.8 \\ 
    58800.9 & \swift\ & 00012172008 & 2.5 & \nustar\ & 80502324004 & 67.7 \\
    58812.3 & \swift\ & 00012172018	& 1.7 & \nustar\ & 80502324006 & 48.6 \\
    \hline\hline
     \end{tabular}
    \caption{A summary of the X-ray observations of the high-flux soft state of \src\ analysed in this work. }
    \label{tab_obs}
\end{table*}

\subsubsection{\nustar\ Data Reduction}

We used the standard pipeline NUPIPELINE, part of HEASOFT V6.33 package, to reduce the \nustar\ data. The \nustar\ calibration version V20231228 was used. We extracted source spectra from circular regions with radii of 120\,arcsec, and the background spectra from nearby circular regions with radii of 150\,arcsec. We used the NUPRODUCTS task for this purpose. The 3-78\,keV band is considered for both FPMA and FPMB spectra. 

\subsubsection{\swift\ Data Reduction}

The \swift\ observations are processed using the standard pipeline XRTPIPELINE of HEASOFT. The calibration file version used is 20230723. The two \swift\ observations have a very high count rate. Therefore, for these two observations, we used the same annulus source extraction region of an inner radius of 20 arcsec and outer radius of 47 arcsec as in previous work \citep{Lazar21}. A background spectrum was also extracted from an active detector region far from the source.

\subsubsection{\nicer\ Data Reduction}

We reduced the \nicer\ observation using the NICERL2 command and the
latest calibration files available in CALDB release 20221001. We selected good time intervals (GTIs)  using NIMAKETIME to filter out intervals when the particle background was high ($\rm{KP}<5$ and $\rm{COR\_SAX}>4$) and avoid times of
extreme optical light loading. We then used the NIEXTRACT-EVENTS tool to apply GTIS to the event lists. Source and background spectra were then generated using
the nibackgen3C50 tool v7 \citep{Remillard22}. We then generated spectral response files using NICERARF and NICERRMF.

When plotted, all of our spectra are grouped to have a minimum number of photon counts of 20 per bin. 
We conducted the spectral analysis using XSPEC \citep{Arnaud96} and implemented $\chi^{2}$ statistics to estimate goodness of fit.




\section{Results}
\subsection{Traditional relativistic disc models}

\begin{figure}
    \centering
    \includegraphics[width=.99\linewidth]{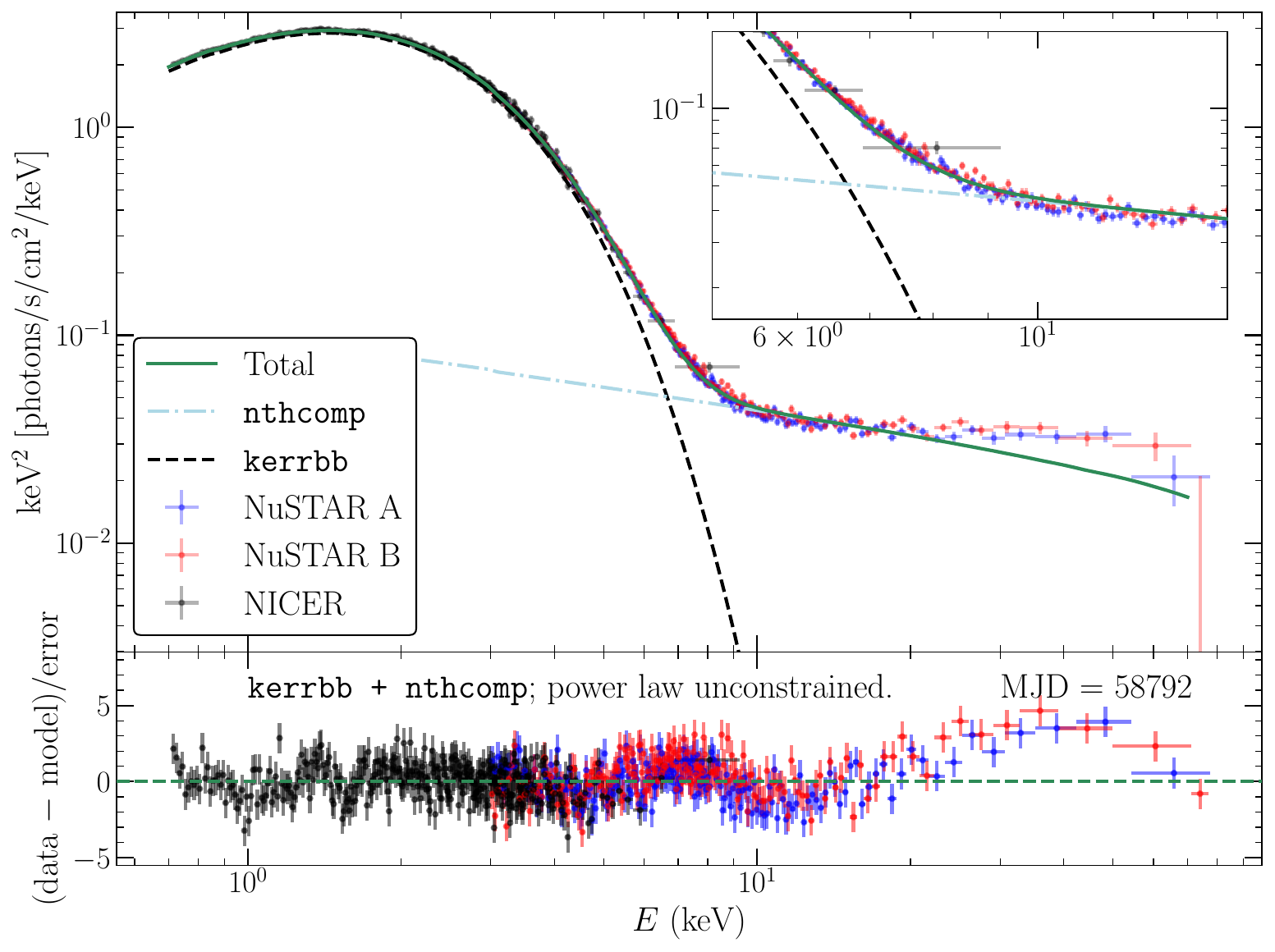}
    \includegraphics[width=.99\linewidth]{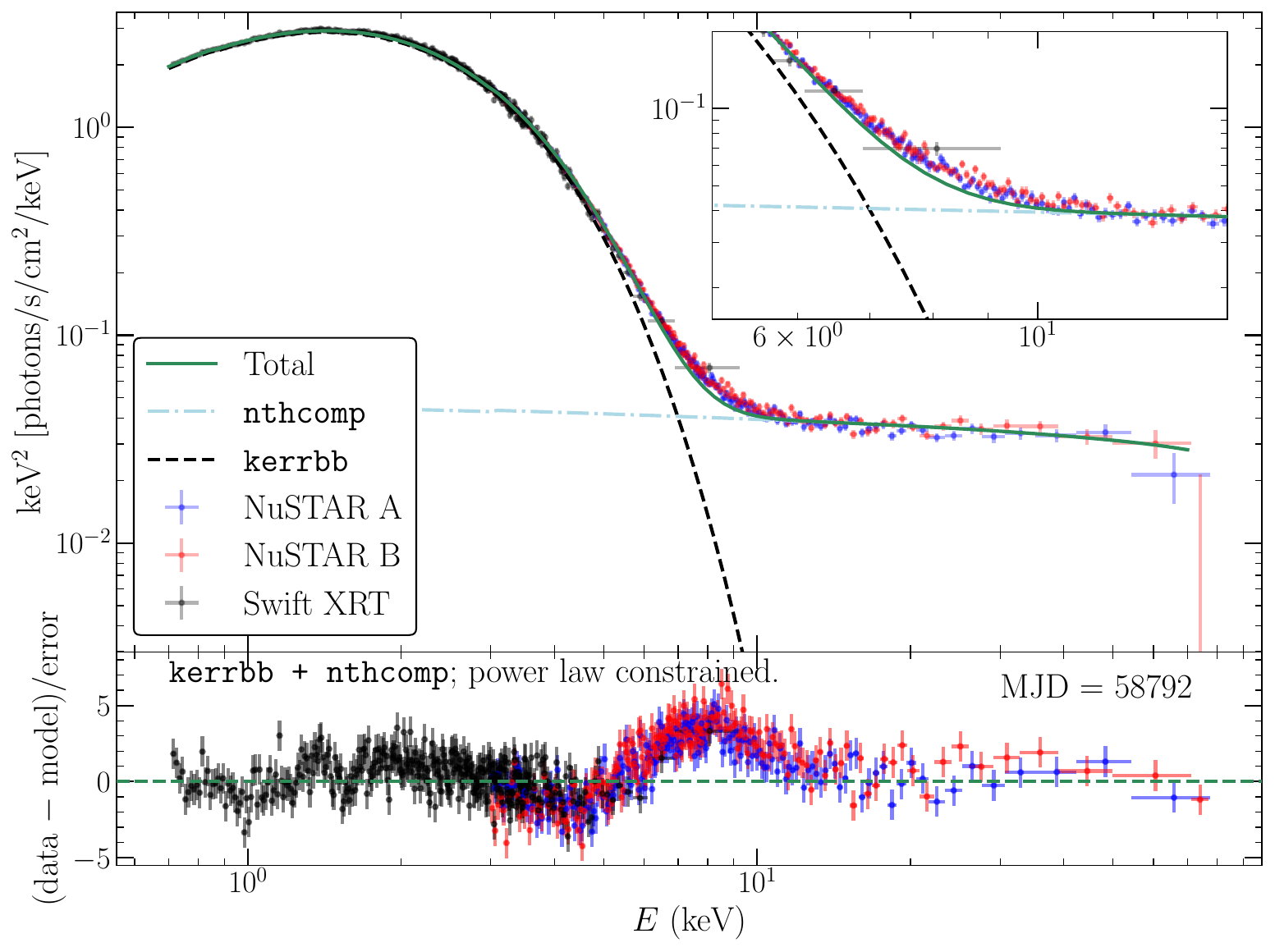}
    \caption{Upper panel: the  best-fitting ${\tt constant} \times {\tt tbabs} \left({\tt kerrbb} + {\tt nthcomp} \right)$ model fit to the  58792 MJD epoch data. It is clear to see that the high photon energy residuals for the formal best-fitting model (upper panel) are a result of the wrong power-law index being fit to the data, a result of the Comptonised component trying to compensate for the insufficient disc flux around $8$ keV. Lower panel: the best-fitting model where the ${\tt nthcomp}$ parameters are first fit to the data at high photon energies $E > 15$ keV, to which they are then frozen. The inset highlights the lack of disc flux at $E \sim 6-10$ keV, the cause of the model-data discrepancy, which can also be seen in the model-data residuals plotted below.  }
    \label{fig:kerrbb-fail2}
\end{figure}

We begin by recovering the results of \cite{Lazar21}, chiefly that the inclusion of high photon energy data from \nustar\ reveals the failure of standard two-component black hole accretion models to reproduce the MAXI J0637$-$430 data. In the syntax of {\tt XSPEC} we fit the model 
\begin{equation}
    {\tt model} = {\tt constant} \times {\tt tbabs} \left({\tt kerrbb} + {\tt nthcomp} \right) , 
\end{equation}
to the combined \swift\ XRT, \nicer\ and \nustar\ data sets. The simple multiplicative factor ${\tt constant}$ is included to model any systematic  differences between the different observatories (and the two different \nustar\ modules)  and was fixed to unity for the \nustar\ A module. For each epoch analysed in this {\it Letter} the best-fitting ${\tt constant}$ values found between the instruments are consistent within the known calibration uncertainty of \nustar\ \cite{Madsen15}.  Absorption of X-ray photons by neutral hydrogen along the line of sight is modelled by the ${\tt tbabs}$ convolution model \cite{Wilms00}, which introduces a hydrogen depth fitting parameter $N_{\rm H}$. Finally, ${\tt nthcomp}$ models the Comptonisation of disc photons on their passage through the disc corona, and is included to model the high energy power-law tail observed from MAXI J0637$-$430 (e.g., Fig. \ref{fig:kerrbb-fail2}). The disc model is ${\tt kerrbb}$ \citep{Li05}, which corresponds physically to the  \cite{NovikovThorne73} relativistic thin disc solution with an imposed vanishing ISCO stress boundary condition. 

As can be seen in Figure \ref{fig:kerrbb-fail2}, this model is unable to reproduce the data \citep[a finding also reported in][]{Lazar21}. {In this fit we allowed the black hole spin, disc-observer inclination angle,  accretion rate and disc colour-correction to freely vary, while fixed the mass and distance to the \cite{Soria22} values. } This {inability to fit the data} remains the case even when the model is able to explore regions of parameter space which are unphysical (e.g., extra-galactic distances and unphysical black hole masses). 

\begin{figure}
    \centering
    \includegraphics[width=.95\linewidth]{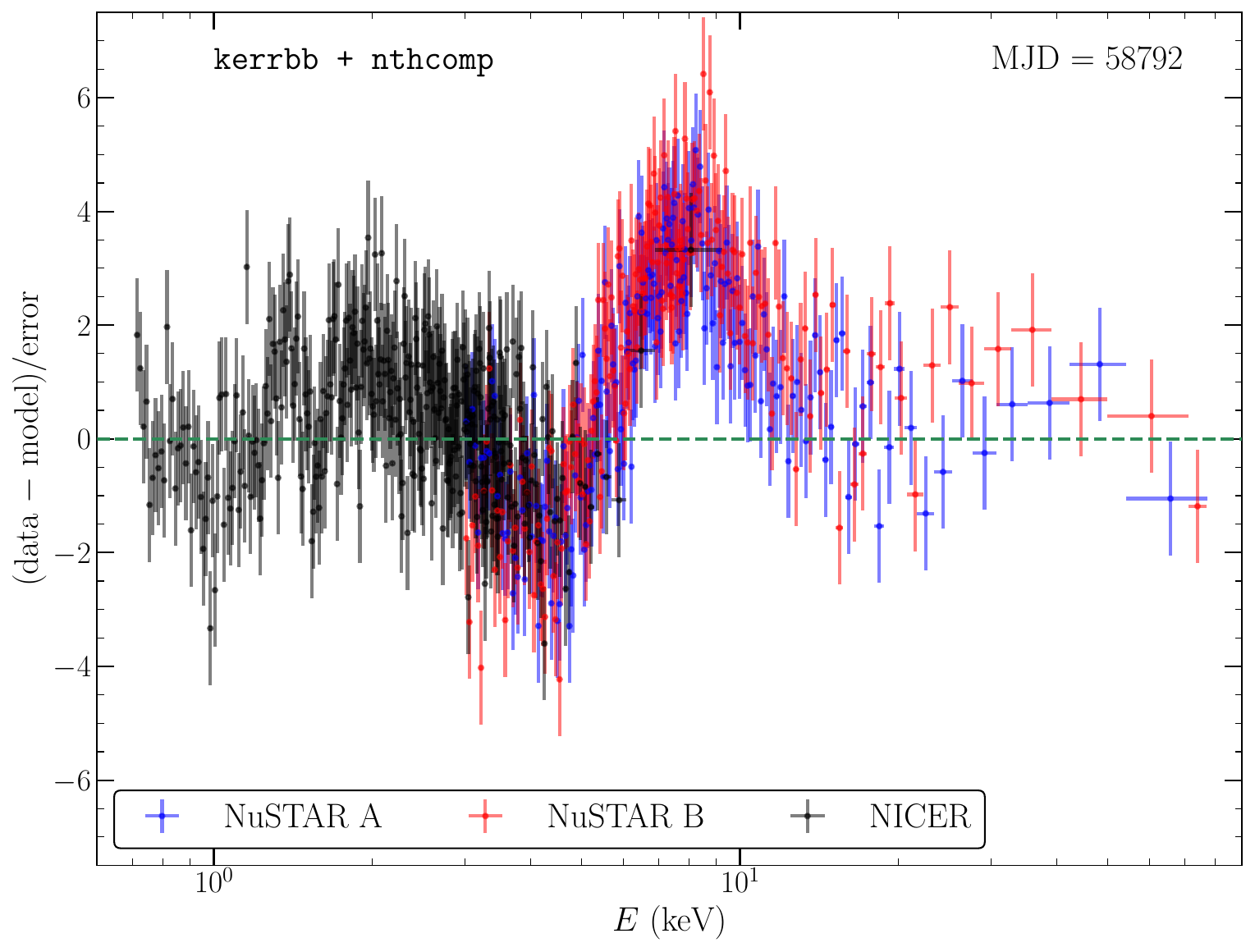}
    \includegraphics[width=.95\linewidth]{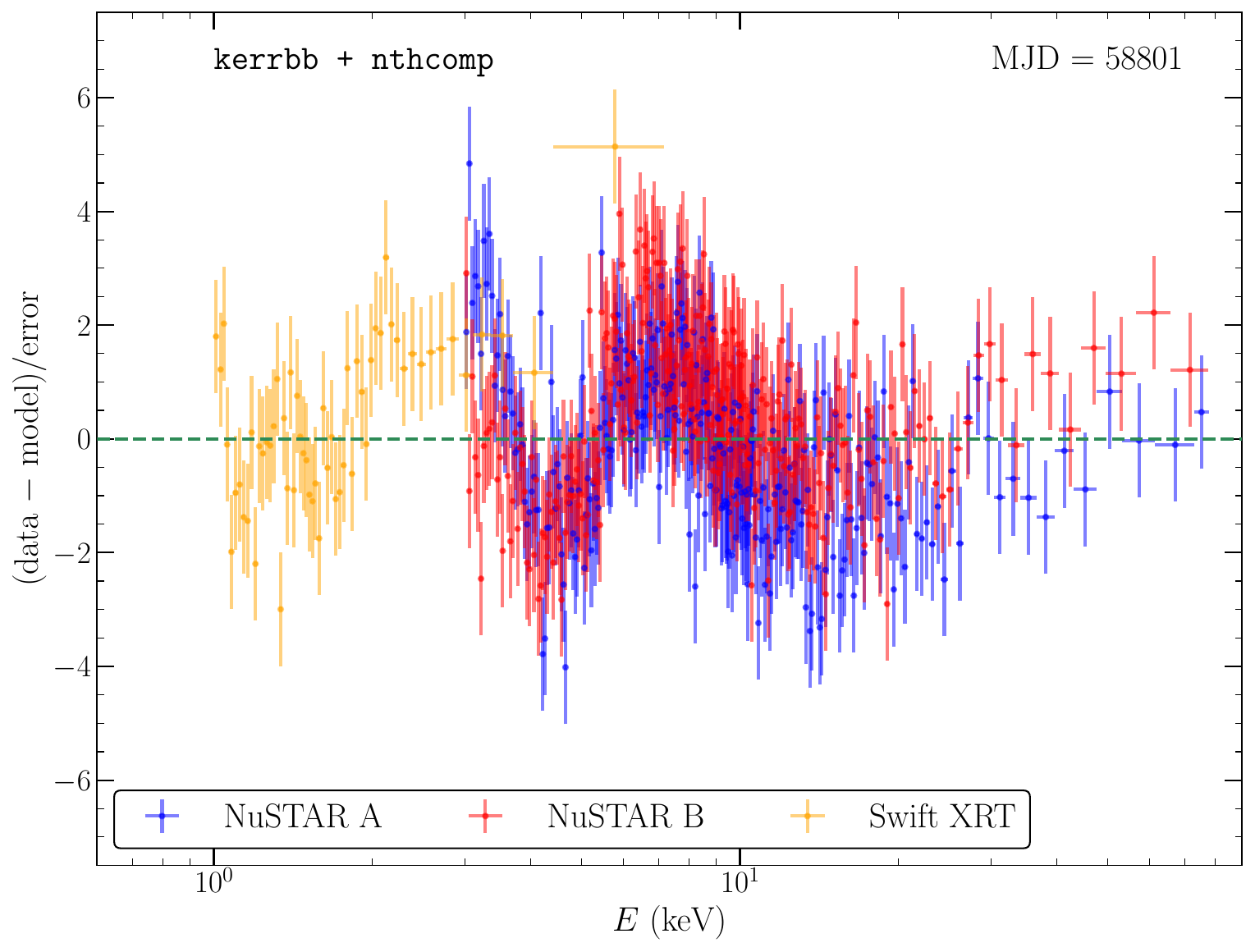}
    \includegraphics[width=.95\linewidth]{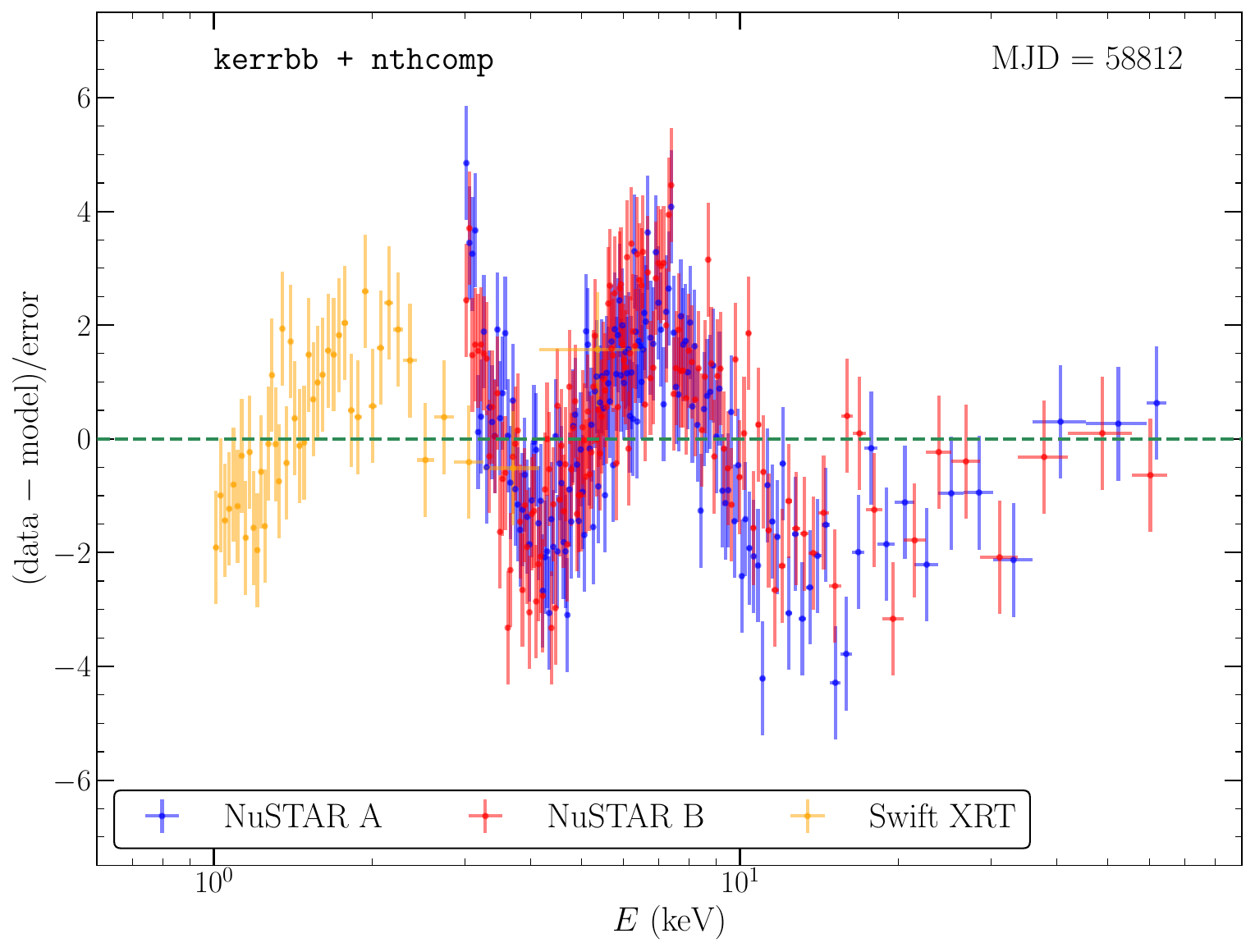}
    \caption{The residuals of the best-fitting ${\tt constant} \times {\tt tbabs} \left({\tt kerrbb} + {\tt nthcomp} \right)$ model to the MAXI J0637$-$430 data. This model is unable to produce an acceptable fit to the data, with large residuals particularly notable centred on $E\sim 8$ keV.  }
    \label{fig:kerrbb-fail1}
\end{figure}

If we allow the power-law index $\Gamma$ of the ${\tt nthcomp}$ model to vary freely during fitting, then the largest residuals between the model and data found at the highest photon energies (Fig. \ref{fig:kerrbb-fail2}, upper panel), with less extreme residuals located around $E \sim 6-10$ keV. This, however, does not accurately reflect the failure points of the model, which is entirely restricted to the region $E \sim 6-10$ keV. This can be most clearly seen in the lower panel of Figure \ref{fig:kerrbb-fail2}, where we show the model fits for the observational epoch 58792 MJD. The upper panel shows the formally best-fitting model to the data, where it is clear to see that the high photon energy residuals are a result of the wrong power-law index being fit to the data, a result of the Comptonised component trying to compensate for the insufficient disc flux around $6$ keV. 

If, instead, we first fit the ${\tt nthcomp}$ parameters to the data at high photon energies $E > 15$ keV, to which they are then frozen before the disc parameters are fit, we see the true location of the model-data discrepancy. This fit is shown in the lower panel, with a zoomed-in inset highlighting the region $E\sim 6-10$ keV, where the disc model produces insufficient flux (note also the residuals in the lowest panel). This behaviour is reproduced in epochs 58801 and 58812.  By adding a third, pure blackbody, component this additional flux can be accurately reproduced \citep[as first shown by][]{Lazar21}, although the physical origin of this additional blackbody is unclear.   This exact signature was seen in another X-ray binary MAXI J1820+070 \citep{Fabian20}, which was later shown to be attributable to intra-ISCO emission \citep{Mummery24Plunge}. 

We repeat this analysis, namely first fitting the ${\tt nthcomp}$ parameters to the data at high photon energies $E > 15$ keV, to which they are then frozen before the disc parameters are fit, to the other two epochs and show the model-data residuals in Figure \ref{fig:kerrbb-fail1}. Every observational epoch shows an insufficient disc flux centred around $E\sim 8$ keV (note that this is higher than the iron line rest frame energy $E = 6.4$ keV).  The key physical question is whether this excess emission can be accurately and self-consistently reproduced with a model which includes emission produced within the plunging region. 



\subsection{Including plunging region emission}
\begin{figure}
    \centering
    \includegraphics[width=\linewidth]{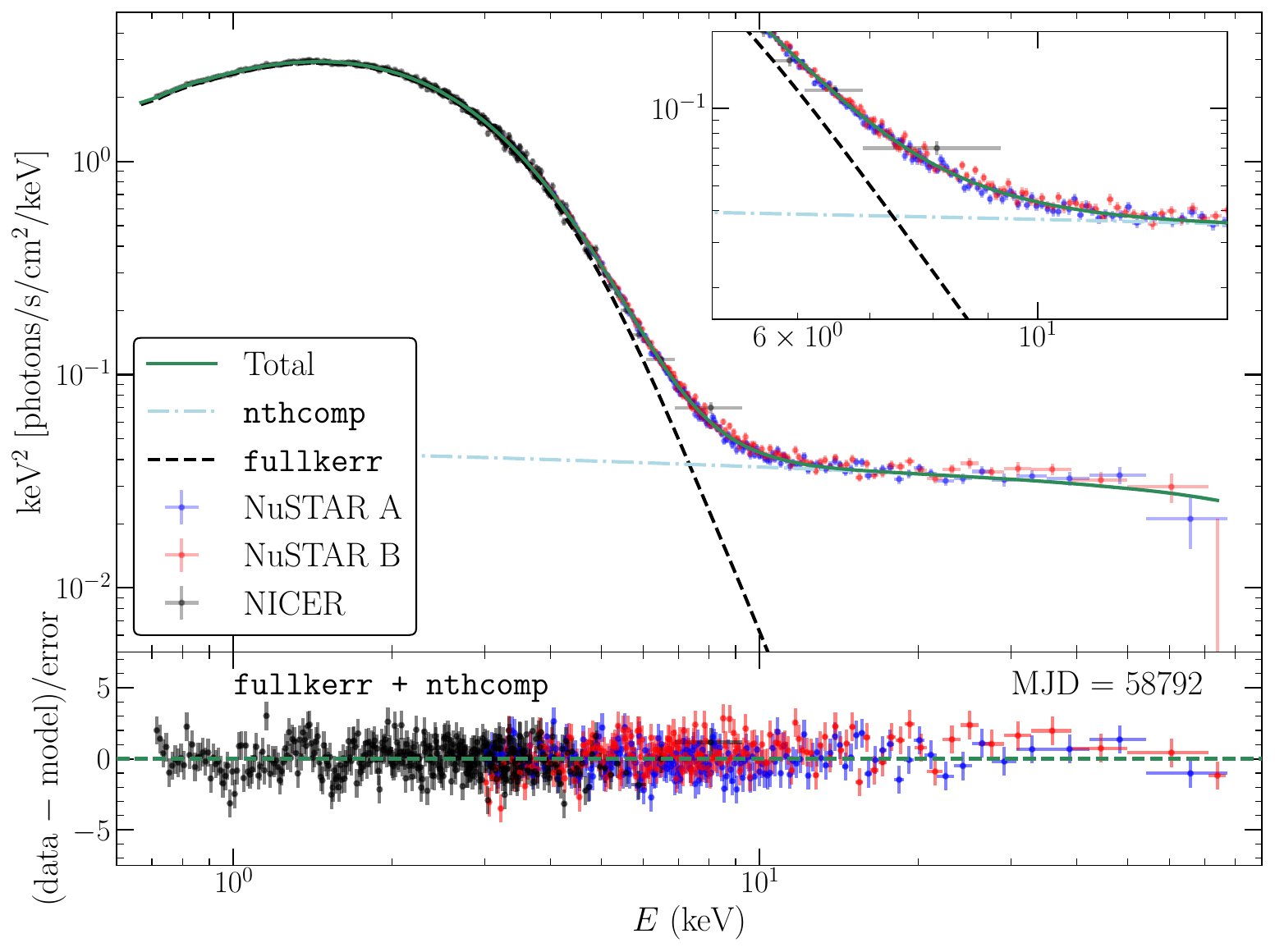}
    \includegraphics[width=\linewidth]{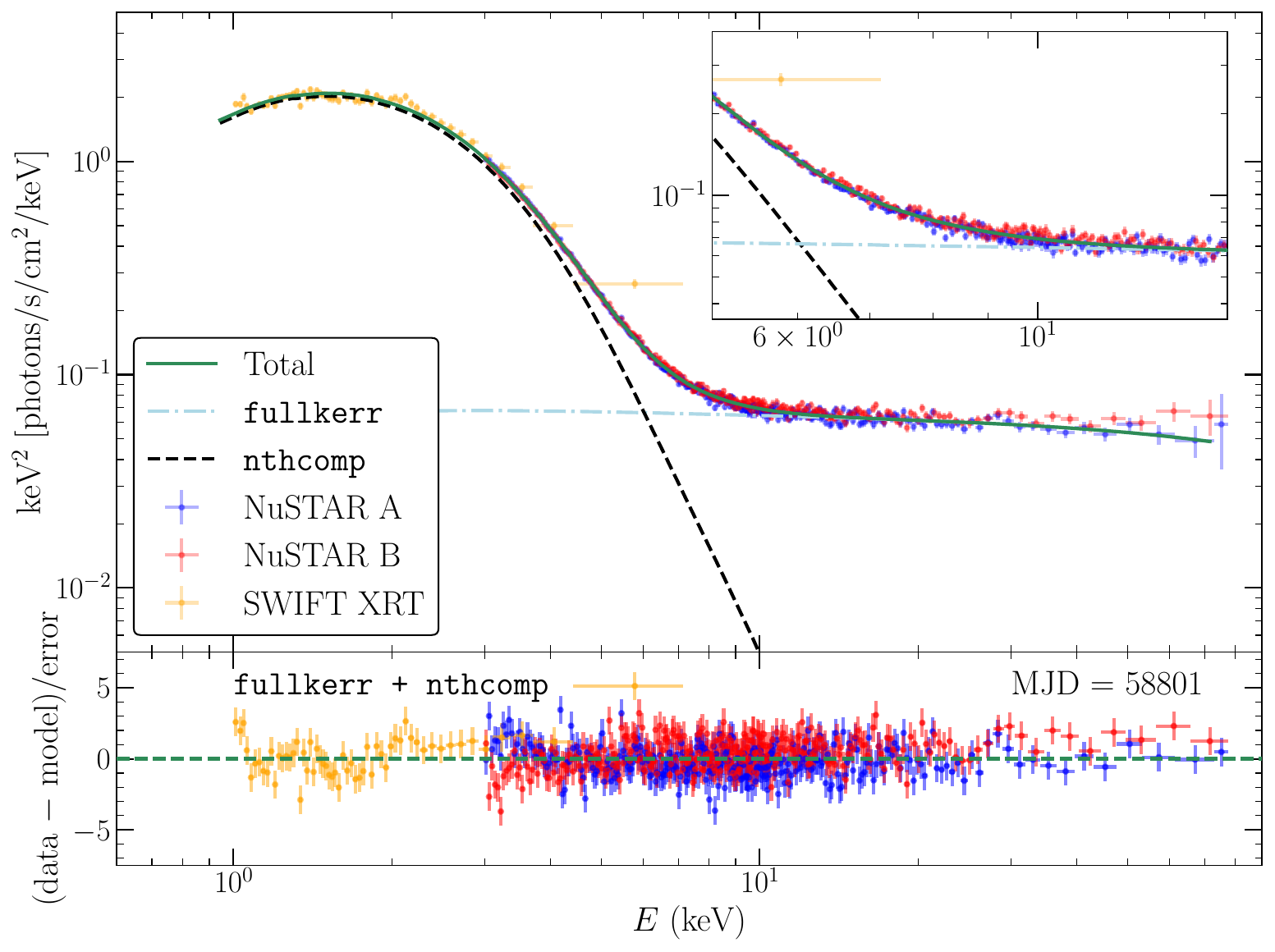}
    \includegraphics[width=\linewidth]{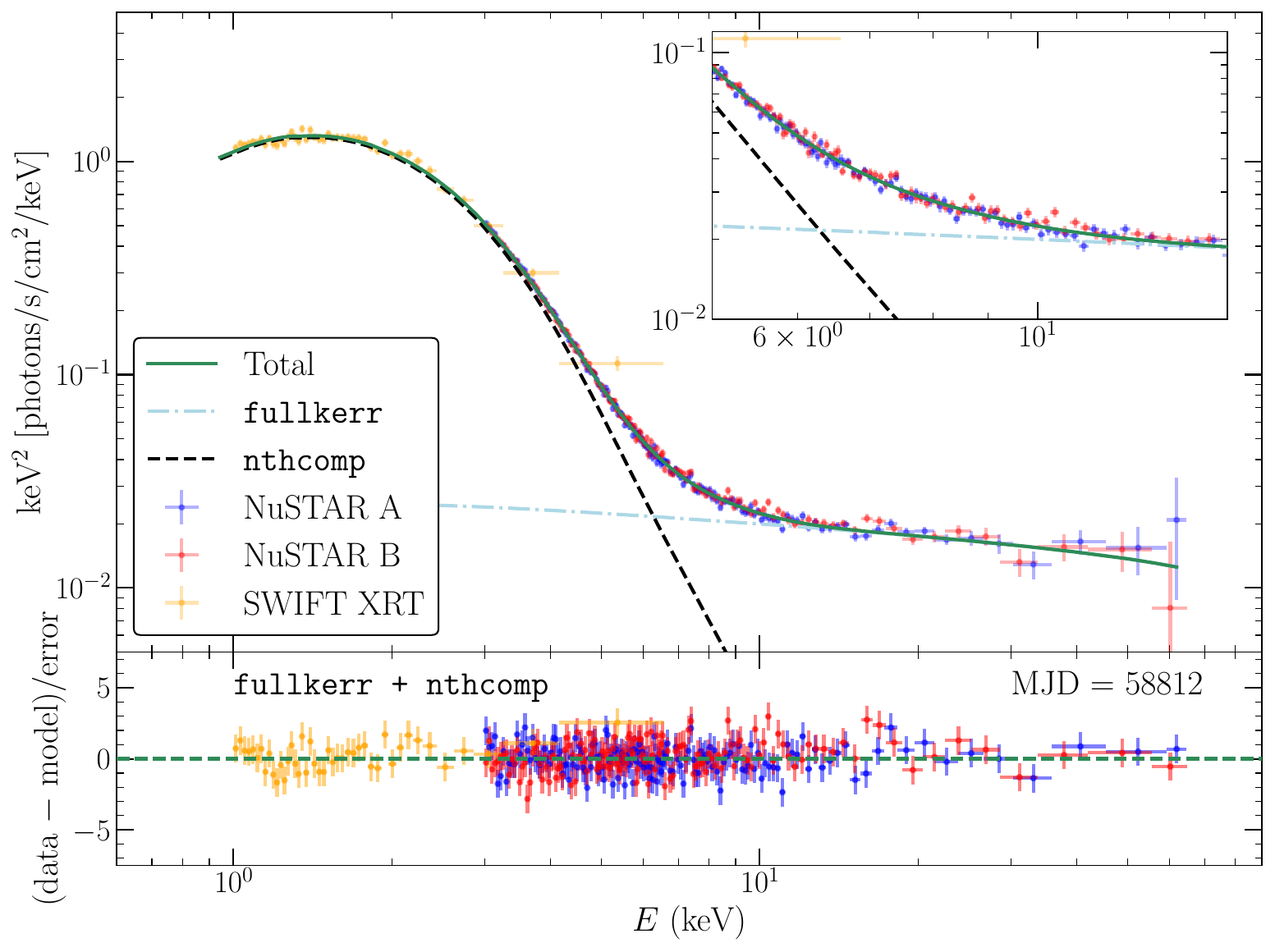}
    \caption{The X-ray spectra of MAXI J0637$-$430 at the three different epochs used in this work (dates displayed on each plot), with the different model components displayed explicitly. The lower panels shows the residuals of the fit, which are significantly reduced with the inclusion of intra-ISCO emission. The inset in each upper panel shows a zoomed in view of the $E = 5-15$ keV spectral region which is problematic for traditional models.   These fits were found with the following parameters frozen $D = 8.7$ kpc, $M_\bullet = 5.1 M_\odot$, $a_\bullet = 0$. For these fixed parameters a disc-observer inclination angle of $i = 17^\circ$ was found to reproduce the data. }
    \label{fig:data+fits}
\end{figure}

Now that we have determined that the X-ray spectrum of MAXI J0637$-$430 cannot be well described by conventional two-component models which neglect intra-ISCO emission \citep[as was previously noted by][]{Lazar21}, we turn to models which include this component. To remain consistent with the previous section we fit (in the syntax of {\tt XSPEC}) the model 
\begin{equation}
    {\tt model} = {\tt constant} \times {\tt tbabs} \left({\tt fullkerr} + {\tt nthcomp} \right) , 
\end{equation}
to the combined \swift\ XRT, \nicer\ and \nustar\ data sets, i.e., the only change in the analysis is the change in disc model from ${\tt kerrbb}$ to ${\tt fullkerr}$. In contrast to ${\tt kerrbb}$ we find that this new model can accurately reproduce the data for a broad range of parameter space.  In particular, there is a broad range of values for the black hole $(M_\bullet, a_\bullet)$ and observer $(D, i)$ parameters which can all reproduce the data. 

In the outer regions of the main body of the disc  \texttt{fullkerr} and \texttt{kerrbb} are functionally identical: they both consistent of the \cite{NovikovThorne73} relativistic disc model, with a simple colour-correction factor $f_d$ which models the effects of radiative transfer in the disc atmosphere \citep[e.g.,][]{Shapiro83}. At the ISCO however,  \texttt{fullkerr} allows for a general boundary condition (unlike the enforced vanishing temperature condition in \texttt{kerrbb}), parameterised by a dimensionless free-parameter $\delta_{\cal J}$. This parameter corresponds physically to the fraction of the disc fluid's angular momentum which is “passed back” to the extra-ISCO accretion flow by magnetic stresses acting within the plunging region \citep[e.g.,][]{MummeryBalbus2023}, the value of which is expected (from simulations), to be at the $\delta_{\cal J} \sim 0.01-0.1$ level \citep{Penna10, Noble10}. Within the ISCO radius (denoted $r_I$) the disc fluid transitions to a ``photon-starved'' state with colour-correction factor which grows as $f = f_d(r/r_I)^{-\xi}$ (see \met, \citealt{Davis19} and \citealt{Zhu12} for further discussion of the physics of the radiative transfer in this region). 


As discussed previously, there are currently no tight constraints on the four  parameters $(M_\bullet, a_\bullet, D, i)$. To perform a concrete analysis we therefore  take the parameters found by \cite{Soria22}, namely $M_\bullet = 5.1 M_\odot$, and $D = 8.7$ kpc. There remains a degeneracy between black hole spin and disc-observer inclination, so for this analysis we fixed the black hole spin to that of a Schwarzschild spacetime $a_\bullet = 0$, finding that $i = 17^\circ$ was sufficient to reproduce the data. The reason we restrict the spin in this manner is that we are interested primarily in the signatures of the plunging region, not on the precise parameters of the black hole itself. We return to the black hole's spin parameter in a later subsection.  With these four parameters fixed, we fit the remaining disc $(\dot M, f_d, \delta_{\cal J}, \xi)$, corona\footnote{Formally ${\tt nthcomp}$ can also model different temperature parameters  for the seed photon field $(T_{\rm bb})$ and scattering electron population $(T_e)$. We found no sensitivity to either of these parameters, provided that $T_e \gtrsim 100$ keV, and so did not let them freely vary.  } $(\Gamma, \, {\tt norm})$ and absorption $N_H$ parameters. The best fitting disc parameters, and $90\%$ confidence intervals, are summarised in Table \ref{epochs}. 



\begin{table}
    \renewcommand{\arraystretch}{2}
    \centering 
    \begin{tabular}{ | p{1.7cm} | p{1.65cm} p{1.65cm} p{1.65cm} | }
    \hline
       \diagbox[width=2.14cm]{Parameter}{Epoch}  & 58792 & 58801 & 58812   \\
    \hline 
       $\dot M$ & $1.36^{+0.0078}_{-0.0087}$ & $1.03^{+0.02}_{-0.02}$ & $0.697^{+0.023}_{-0.022}$ \\
       $\delta_{\cal J}$ & $0.0294^{+0.0016}_{-0.0014}$ & $0.0354^{+0.0033}_{-0.0027}$ & $0.0344^{+0.0034}_{-0.0026}$  \\ 
       $\xi$ & $ 2.05^{+0.11}_{-0.11}$ & $2.19^{+0.14}_{-0.14}$ &  $2.41^{+0.20}_{-0.19}$ \\ 
       $f_d$ & $1.58^{+0.012}_{-0.013}$ & $1.52^{+0.037}_{-0.040}$ & $1.51^{+0.047}_{-0.055}$\\ 
       \hline 
       $\chi_{r, {\tt fk}}^2$ &  $\frac{1861.5}{1716}$ & $\frac{2050.8}{1906}$& $\frac{1225.8}{1237}$ \\ 
       $\Delta \chi_{{\tt fk} - {\tt kbb}}^2$ & $-1338.3$ & $-632.6$ & $-707.4$ \\
       \hline
    \end{tabular}
    \caption{Best fitting parameters of the \fk model to each MAXI J0637$-$430 epoch, the  fit statistic of this model and the improvement in the fit statistic compared to a {\tt kerrbb + nthcomp} model. These fits were found with the following parameters frozen $D = 8.7$ kpc, $M_\bullet = 5.1 M_\odot$, $a_\bullet = 0$,  $i = 17^\circ$. Parameters are presented in the units of the \fk model.  Parameter confidence intervals are $90\%$ intervals. The luminosity Eddington ratio is roughly 10$\%$ of the reported accretion rate parameter.   }
    \label{epochs}
\end{table}

The results of this fitting procedure are shown in Figure \ref{fig:data+fits}. In the upper panels we display the X-ray spectra of MAXI J0637$-$430 at the three different epochs used in this work (dates displayed in the upper right corner of each plot), with the different model components (${\tt fullkerr}$ and ${\tt nthcomp}$) displayed explicitly. The lower panels show the residuals of the fit at each epoch, while the zoomed in region displays the $E = 5-15$ keV spectral region which is problematic for traditional models (Fig. \ref{fig:kerrbb-fail1}).  A visual inspection of the differing behaviour of the residuals of the ${\tt fullkerr}$ model (Fig. \ref{fig:data+fits}) and ${\tt kerrbb}$ model (Fig. \ref{fig:kerrbb-fail1}) highlights the improvement in the quality of the fit to the data by including intra-ISCO emission. The formal improvement in chi-squared fit statistic is $\Delta \chi^2 = - 1338.3$ (epoch 58792), $\Delta \chi^2 = - 632.6$ (epoch 58801) and $\Delta \chi^2 = - 707.4$ (epoch 58812)\footnote{If the power-law index $\Gamma$ of the ${\tt nthcomp}$ model is allowed to freely vary then the formal improvement in chi-squared fit statistic is $\Delta \chi^2 = - 267.2$ (epoch 58792), $\Delta \chi^2 = - 300.0$ (epoch 58801) and $\Delta \chi^2 = - 228.2$ (epoch 58812). This remains highly significant. }.  Improvements in chi-squared fit statistic greater than 100 are highly significant, and intra-ISCO emission is required at a high confidence level. The ISCO stress parameter $\delta_{\cal J}$ is found to be $\delta_{\cal J} \simeq 0.03$ in this work, this is within the range found in general relativistic magnetohydrodynamic simulations $\delta_{\cal J} \sim 0.01-0.1$ \citep{Penna10, Noble10}, and very similar to the value found for MAXI J1820+070 \met.

\subsection{Black hole spin constraints}
\begin{figure}
    \centering
    \includegraphics[width=\linewidth]{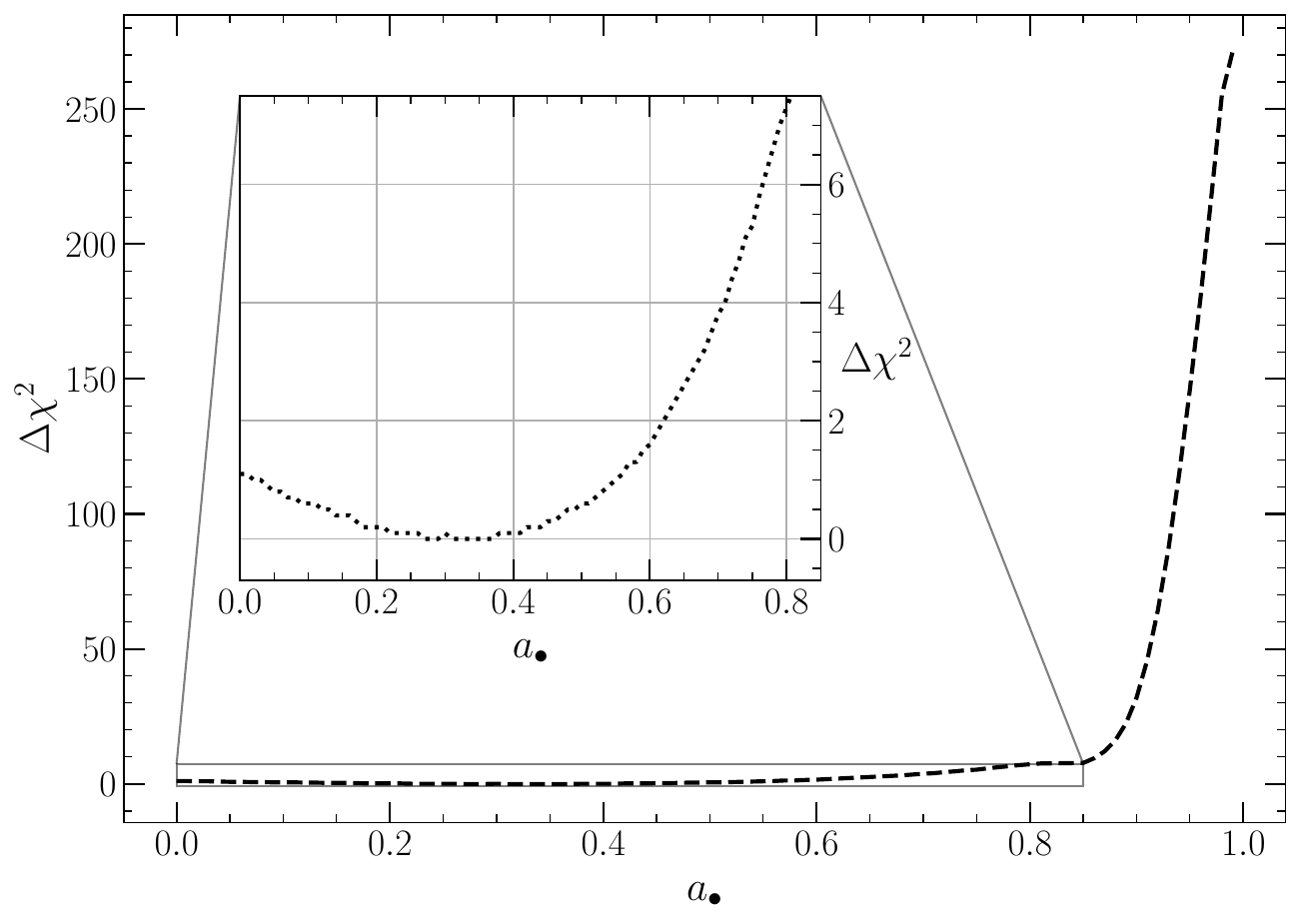}
    \includegraphics[width=\linewidth]{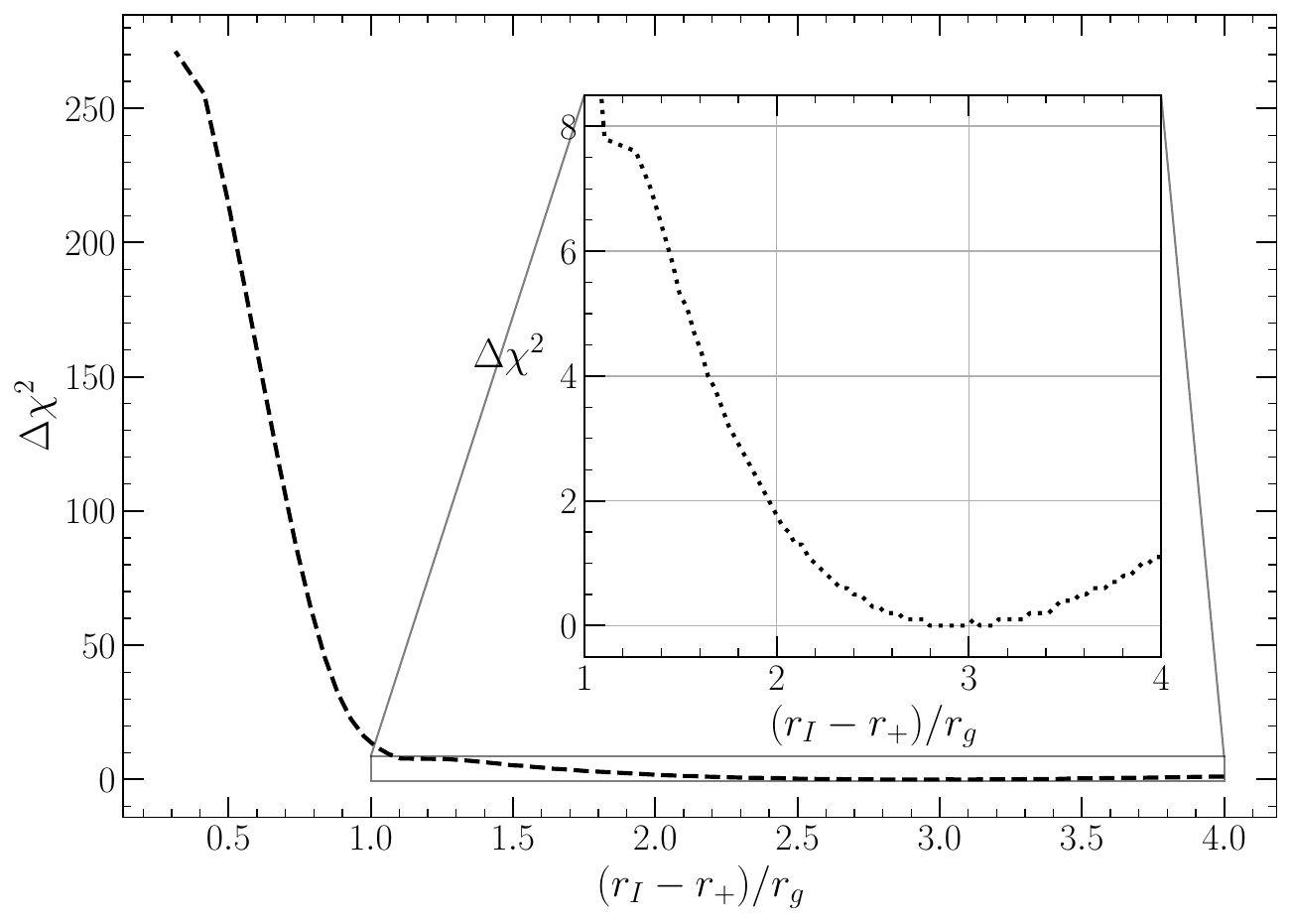}
    \caption{The change in $\chi^2$ fit statistic as a function of black hole spin (upper panel), or equivalently plotted against plunging region size (lower panel). If the  \src\ spin is too high $a_\bullet \gtrsim 0.9$, there is insufficient disc area to power the observed $E\sim 8$ keV intra-ISCO component, and the fit rapidly becomes  poor. The two insets display a region of parameter space which is formally acceptable.  }
    \label{fig:spin_constraints}
\end{figure}
In this work we have uncovered a robust detection of photons emitted from within the plunging region of the X-ray binary \src. Obviously, for this to be possible, the plunging region in \src\ must have a non-zero radial extent. An extremal black hole (i.e., one with maximal spin $a_\bullet = 1$) has no observable plunging region, as the event horizon $r_+(a_\bullet)$ and ISCO radius $r_I(a_\bullet)$ coincide $r_+ = r_I = r_g = GM_\bullet/c^2$. 
This simple fact implies that we can place a constraint on the maximum spin of the black hole at the centre of \src. We repeat the analysis of the previous section, by refitting the {\tt fullkerr + nthcomp} model to the \nustar, \nicer, and \swift\ data for a range of different spin parameters, recording the minimum chi squared fit statistic (which occurred for a black hole spin parameter $a_\bullet = 0.3$).  In Figure \ref{fig:spin_constraints} we present the difference in chi-squared statistic as a function of black hole spin (upper), or equivalently plunging region size (lower; note that there is a one to one relationship between spin and plunging region size, they are not independent parameters). If the  \src\ spin is too high $a_\bullet \gtrsim 0.9$, there is insufficient disc area to power the observed $E\sim 8$ keV intra-ISCO component, and the fit rapidly becomes poor. We can constrain the black hole spin in \src\ to be $a_\bullet < 0.86$ at the 99.9\% confidence level.

\section{Conclusions} 
In this {\it Letter} we have demonstrated that the soft X-ray spectra of the source MAXI J0637$-$430 can be well described with simple two-component (disc plus Comptonisation) models, provided that emission sourced from within the ISCO is included in the disc model. The improvement, in fit statistic terms, from including intra-ISCO emission is highly significant $(\Delta \chi^2 < -600)$ for all three epochs with high quality data.    If the vanishing ISCO stress boundary condition is enforced then we recover results previously published in the literature \citep{Lazar21} that three component models are required to reproduce the data.   This behaviour is qualitatively identical to that seen in the X-ray binary MAXI J1820+070, the only other currently known system with intra-ISCO emission \citep{Mummery24Plunge}. 

\begin{figure}
    \centering
    \includegraphics[width=\linewidth]{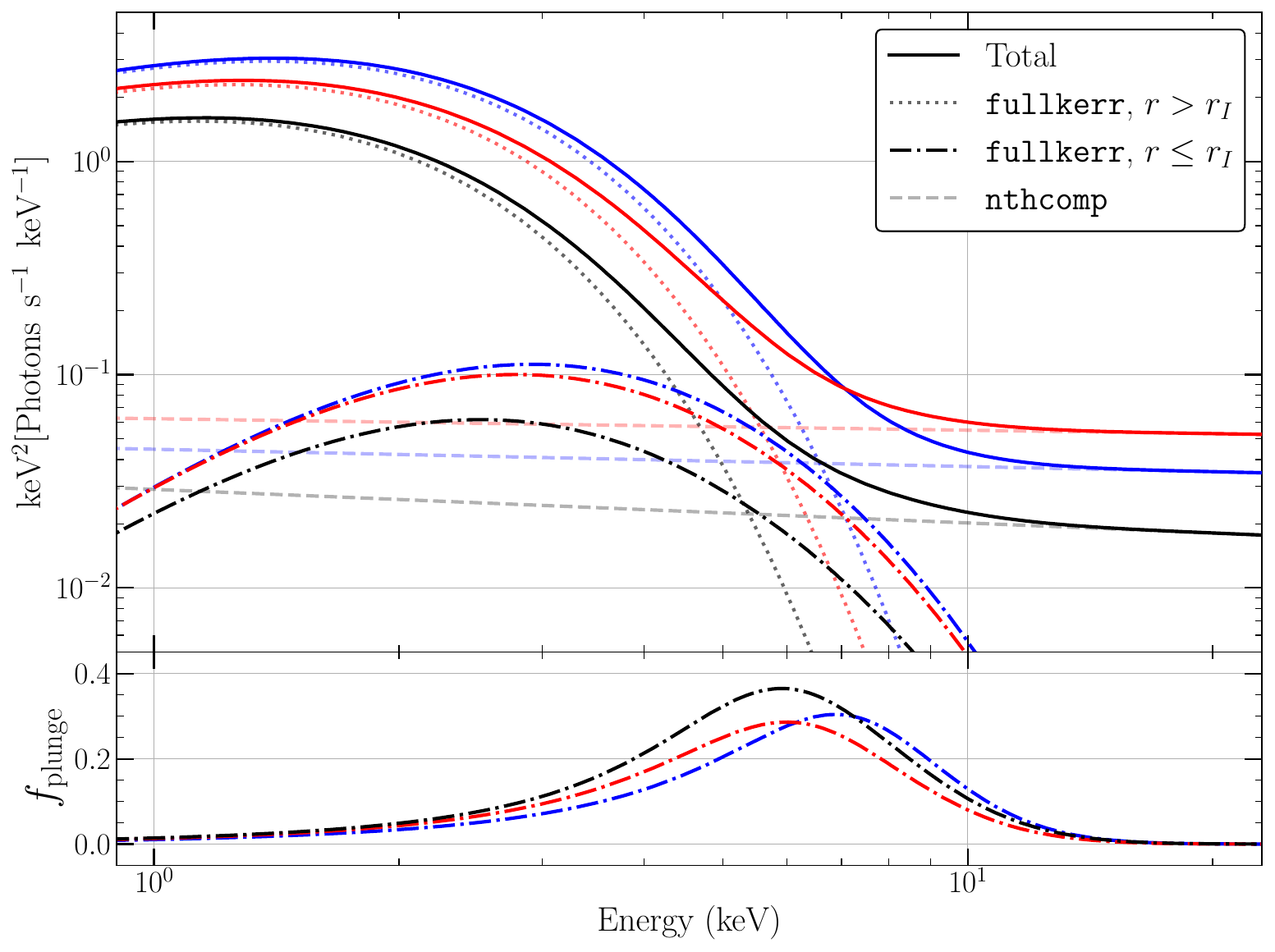}
    \caption{Upper panel: the individual contributions of different model components to the total observed flux. Each epoch is denoted by a different colour, with the earliest epoch MJD = 58792 distinguishable by the brightest peak flux. Later epochs get progressively fainter. The contribution from the plunging region is denoted by dot-dashed curves, and has the form of a hot-and-small blackbody component. In the lower panel we show the fractional contribution to total flux from radii within the ISCO $f_{\rm plunge} \equiv F_I/F_{\rm tot}$, as a function of observing energy. While relatively insignificant for both soft $E< 2$ keV, and hard $E>20$ keV X-ray bands, at intermediate energies $E \sim 6-10$ keV the flux sourced from within the plunging region represents a significant contribution  to the total observed flux (around $30\%-40\%$). } 
    \label{fig:cont+spec}
\end{figure}

In Figure \ref{fig:cont+spec} we show the contribution of the flux sourced from within the plunging region to the total observed flux, for each of the three epochs with \nustar\ data. While relatively insignificant for both soft $E< 3$ keV, and hard $E>20$ keV X-ray bands, at intermediate energies $E \sim 6-10$ keV the flux sourced from within the plunging region (denoted $F_I$) represents a significant contribution to the total observed flux (denoted $F_{\rm tot}$). The lower panel of Figure \ref{fig:cont+spec} shows this ratio as a function of observing energy, and we find that each epoch has maximum contributions at $E \sim 6$ keV of  $f_{\rm plunge} \equiv F_I/F_{\rm tot} \sim 0.3$. Clearly, the observational features of the plunging region are somewhat subtle, and correspond to a slight flux excess over and above vanishing ISCO stress disc models detectable at around $\sim 6-10$ keV (see inset of Figure \ref{fig:kerrbb-fail2}).  High quality observations are required to disentangle this slight flux excess from other complicating components, and we encourage future dedicated observational probes of this spectral region.

In addition, in this paper we have also highlighted the necessity of including high photon energy $(E\gtrsim 20$ keV$)$ data in continuum fitting exercises. As highlighted by Figure \ref{fig:kerrbb-fail2}, it is essential that these high energy observations are included so that the signatures of the plunging region are not modelled out by coronal components with the wrong power-law indices which can, over a restricted interval, successfully mimic the additional photon flux sourced from within the ISCO. \nustar\ may therefore be the best instrument currently available for probing this region, although we note that Insight-HXMT, historic data taken by RXTE, and in the future HEX-P \citep{Madsen24},  all cover this broad region. 

{With a robust detection of intra-ISCO emission, we are able to place a constraint on the {\it maximum} angular momentum which the black hole in MAXI J0637$-$430 can have. The physical reason for this is simple, if the  \src\ spin is too high $a_\bullet \gtrsim 0.9$, the size of the plunging region shrinks (having zero size for a formal $a_\bullet = 1$ system), and there is insufficient disc area to power the observed $E\sim 8$ keV intra-ISCO component, and the fit rapidly becomes poor. We can constrain the black hole spin in \src\ to be $a_\bullet < 0.86$ at the 99.9\% confidence level. The ability to place upper bounds on black hole spins using intra-ISCO emission is an important result, as it compliments but is distinct from existing techniques such as reflection spectroscopy \citep[e.g.][]{Reynolds13}.  Future population studies determining the relative fraction of X-ray binary systems with detectable intra-ISCO emission is of real interest, as this can be leveraged as a probe of the number of systems with black hole's with at most moderate spins.  }

It appears to the authors that it is likely that signatures of the plunging region are more widespread in the spectra of X-ray binaries than has been previously realised, particularly with the second such source now discovered.  We expect that the detection and detailed analysis of this region likely has much to tell us about the physical properties of the black holes at the heart of these systems.


\section*{Acknowledgments}
 AM would like to acknowledge the support of Adam Ingram, Shane Davis and Steven Balbus in creating the \fk model. 
 This work was supported by a Leverhulme Trust International Professorship grant [number LIP-202-014]. J.J. acknowledges support from the Leverhulme Trust, Isaac Newton Trust and St Edmund's College. For the purpose of Open Access, AM has applied a CC BY public copyright licence to any Author Accepted Manuscript version arising from this submission. This research has made use of data from the \nustar\ mission, a project led by the California Institute of Technology, managed by the Jet Propulsion Laboratory, and funded by the National Aeronautics and Space Administration. Data analysis was performed using the \nustar\ Data Analysis Software (NuSTARDAS), jointly developed by the ASI Science Data Center (SSDC, Italy) and the California Institute of Technology (USA). We acknowledge the use of public data from the \swift\ and \nicer\ data archive.

\section*{Data availability}
The {\tt XSPEC} model {\tt fullkerr} is available at the following GitHub repository:  \url{https://github.com/andymummeryastro/fullkerr}. The three X-ray spectra of MAXI J0637$-$430 are all publicly available. All of the data used in this work can be downloaded from the HEASARC website \url{https://heasarc.gsfc.nasa.gov}.

\bibliographystyle{mnras}
\bibliography{andy}

\label{lastpage}

\end{document}